\documentclass[conference]{IEEEtran}
\IEEEoverridecommandlockouts
\usepackage{cite}
\usepackage{amsmath,amssymb,amsfonts}
\usepackage{algorithmic}
\usepackage{graphicx}
\usepackage{textcomp}
\usepackage{xcolor}
\usepackage{comment}
\usepackage[normalem]{ulem}
\useunder{\uline}{\ul}{}

\def\BibTeX{{\rm B\kern-.05em{\sc i\kern-.025em b}\kern-.08em
    T\kern-.1667em\lower.7ex\hbox{E}\kern-.125emX}}
\begin{document}
\title{Demo: An End-to-End Assurance Framework for AI/ML Workloads in Datacenters}

\author{
    \IEEEauthorblockN{Jit Gupta, Tarun Banka, Rahul Gupta, Mithun Dharmaraj, Jasleen Kaur}
    \IEEEauthorblockA{[$gjit$,$tbanka$,$rahugupta$,$mdharmaraj$,$jkaur$]@juniper.net}
    \IEEEauthorblockA{Juniper Networks, Sunnyvale, CA 94089, USA}
    }

\maketitle

\begin{abstract}
Modern machine learning workloads such as large language model training, fine-tuning jobs are highly distributed and span across hundreds of systems with multiple GPUs. Job completion time for these workloads is the artifact of the application, compute, network and storage performance. In case of failure or degraded performance it is imperative to understand the root cause and possible remediation for the problem for end-to-end assurance. This demo showcases SaaS-based observability and automated troubleshooting for AI/ML workload performance issues using cross-layer telemetry and logs (e.g., Application telemetry, Collective communication logs, GPU Health metrics, Network Flow Data, NIC ROCEv2 telemetry). Different use cases are demonstrated for end-to-end assurance such as Cross-layer Dependency Graph, Cross-layer Service Level Expectations, Automated Root Cause Analysis, GPU-to-GPU application path tracing.

\end{abstract}

\begin{IEEEkeywords}
assurance, AI/ML, SLE, flow analysis, root cause analysis, end-to-end, network, GPU
\end{IEEEkeywords}

\vspace{-5pt}
\section{Introduction}
Distributed machine learning applications, need multiple GPUs due to large model and dataset sizes~\cite{10.1145/3377454}. These applications run across multiple systems with GPUs in on-prem datacenters, communicating over low-latency, high-bandwidth networks like Infiniband or RoCEv2~\cite{10.1145/2934872.2934908}. Communication is managed by collective communication libraries like Nvidia’s NCCL~\cite{weingram2023xccl}. Additionally, the coexistence of multiple workloads necessitates assurance for individual applications.

Assurance for these workloads include cross-layer observability from application layer to underlying compute and network layers.  Some key capabilities under assurance are 1. Generation of cross-layer dependency graphs to determine the real-time dependency of these workloads on the underlying compute and network resources.  2. Continuous Service Level Expectations (SLE) monitoring that determines the health of each layer over short and long-term 3. Anomaly Detection and Root Cause Analysis for automated troubleshooting and remediation that helps identify the possible bottleneck responsible for degraded application workload performance. 4. Application-aware GPU-to-GPU network path tracing to determine end-to-end (E2E) path of the workload traffic between two communicating GPUs for pin-pointing any network bottlenecks for application performance degradation. The demo will showcase effectiveness of the aforementioned capabilities for E2E assurance for multiple  different workloads that can co-exist within the AI/ML clusters. To the best of our knowledge, this is the first E2E assurance framework for machine learning workloads in the datacenter.

\vspace{-5pt}
\section{End-to-End Architecture}
Figure 1 shows the E2E architecture, where each workload is mapped to the hosts and corresponding GPUs. In addition, these are further mapped to the underlying network topology (including NICs and switches). This mapping is determined at run-time by aggregating different types of telemetry such as logs, time-series, topology data from multiple sources across different layers from application to network.  Telemetry and log data from multiple sources such as GPUs, NICs, Switches are used to build the E2E application traffic path. A custom NCCL’s logging functionality in conjunction with Flow Telemetry from the switches helps determine E2E path. In addition this helps in mapping network telemetry to individual applications, and provides visibility to collective communication operation behavior for different AI/ML workloads.

\begin{figure}
  \includegraphics[width=\linewidth]{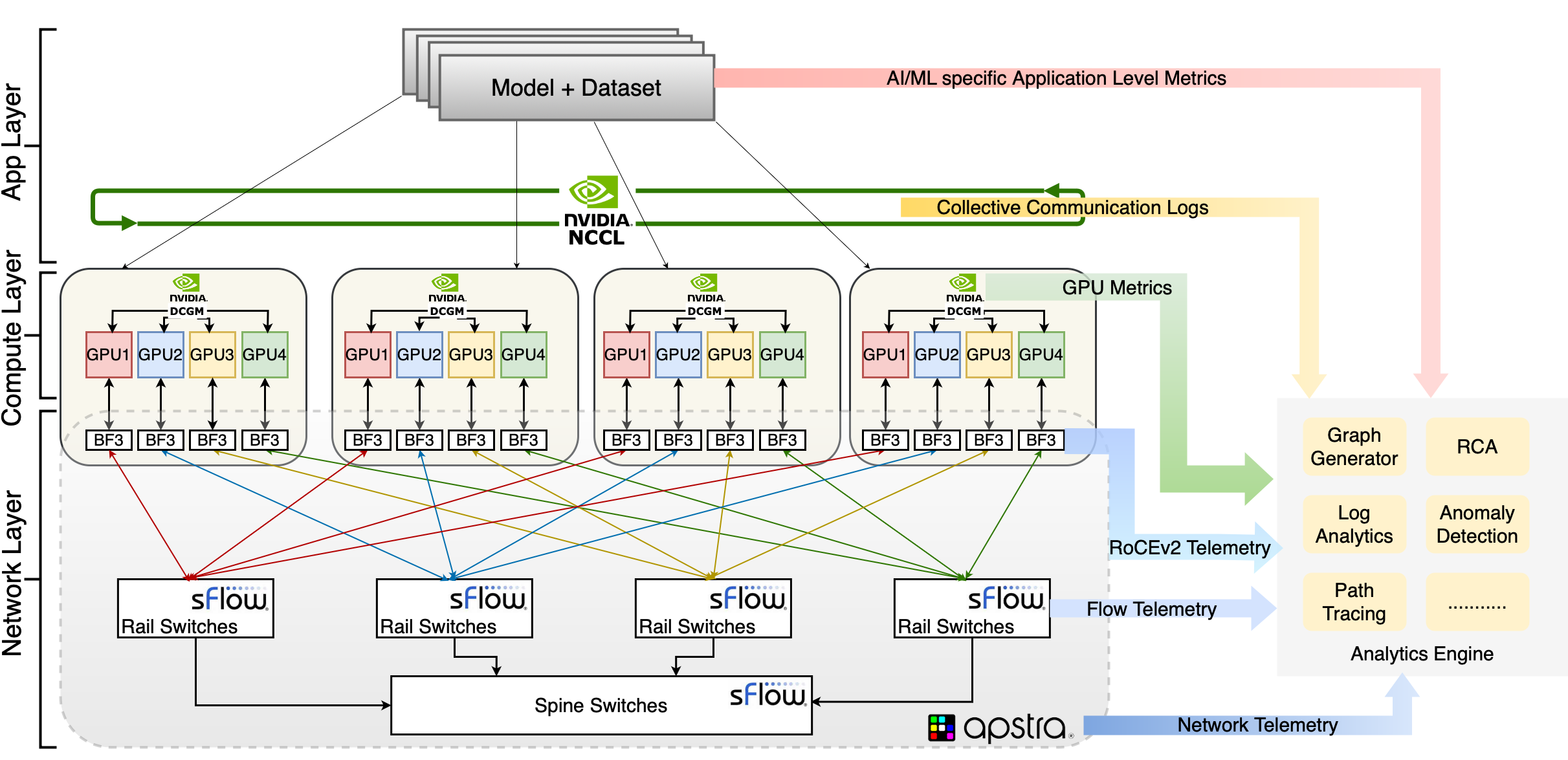}
  \vspace{-15pt}
  \caption{E2E Framework}
  \label{fig:e2eframework}
  \vspace{-15pt}
\end{figure}

The telemetry/metrics from each layer can be classified as:
\begin{itemize}
    \item{Application metrics –} TorchMetrics~\cite{torchmetrics} provides application metrics (ex. iteration rate, model accuracy/loss, etc.). Additionally, we derive operation rate (ex.rate of AllReduce) via our instrumented NCCL logging functionality.
    \item{GPU Metrics:} Nvidia’s Datacenter GPU Manager~\cite{nvidia_dcgm} exposes GPU telemetry such as utilization, energy, etc. 
    \item{NIC Counters–} Hardware counters (ex. detected congestion packets) provide information to identify if RoCE-specific mechanisms are causing network bottlenecks. 
    \item{Flow and Network Telemetry –} Flow level and network telemetry includes sampled packet information, switch interface stats, queue pair counters, etc. to help identify other network bottlenecks.
\end{itemize}

The plug and play nature of the framework allows any telemetry source to be integrated into the assurance piece. For example, NVidia's DCGM can be substituted with a different GPU monitoring tool too. Additionally, this is a SaaS-based solution and hence the data is processed and analyzed in the cloud. Hence it does not interfere with the workload data path.

\begin{figure}
\centering
  \includegraphics[width=0.7\linewidth]{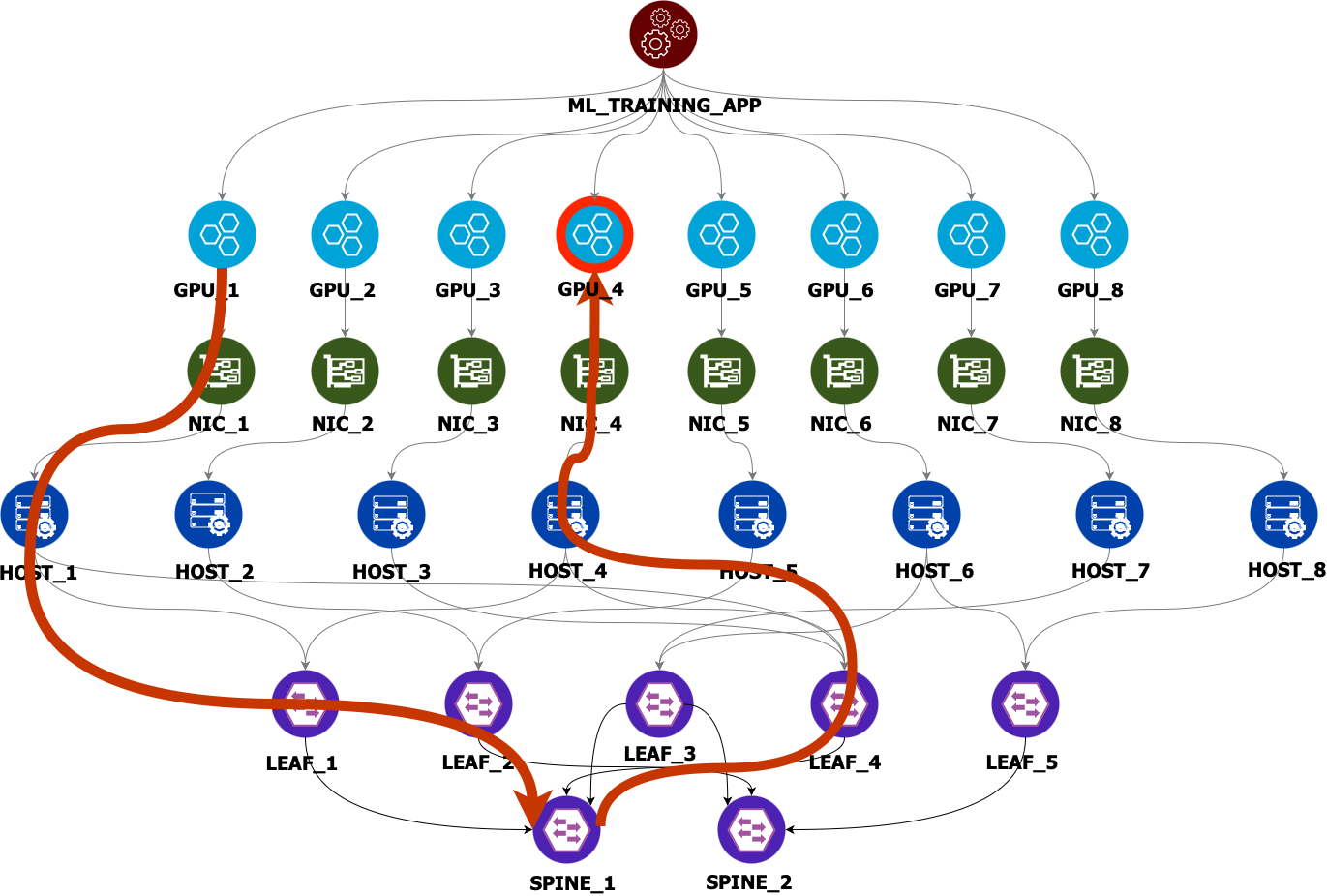}
  \caption{Cross-Layer Dependency Graph}
  \label{fig:uisample}
\end{figure}

\section{Demonstration}
We build a dependency graph between the different layers (similar to  Fig.\ref{fig:uisample}) by discovering the relationship between layer-wise telemetry data. We utilize a setup comprising eight nodes, equipped with an RTX A6000 NVIDIA GPU, connected to Mellanox CX7 NICs with a Clos based network topology. Our instrumented collective communications framework and the torchmetrics module is used to pin a particular application to the set of GPUs that are used to run this application. We do this using the “GPU UUID” label. From NIC level metrics we are able to map individual NICs to hosts, using the “hostname” label. The application used in this overview is a BERT pretraining workload, however we have also experimented with other language, vision and recommendation-based models. Also, the source/destination fields in the flow telemetry helps construct the path taken in the fabric. QP ID extracted from the flows is mapped to NCCL logs associated with an application. This helps to construct the path from the host GPU to remote GPU via the network.

\begin{figure}
    \centering
    \includegraphics[width=0.9\linewidth]{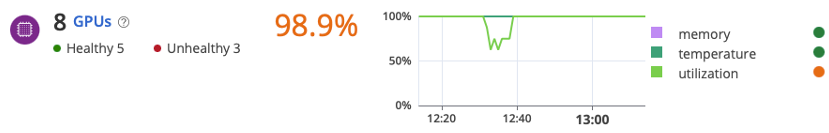}
    \caption{Cross-Layer Service Level Expectations}
    \label{fig:slel}
    \vspace{-15pt}
\end{figure}

\begin{figure}
    \centering
    \includegraphics[width=\linewidth]{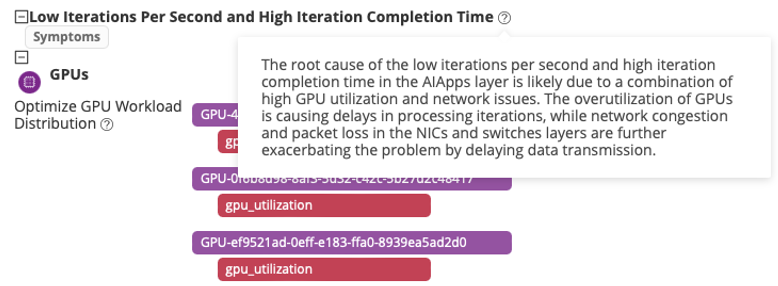}
    \caption{RCA in Action}
    \label{fig:rca}
    \vspace{-15pt}
\end{figure}

We monitor SLE, defined as the percentage of time in a given monitoring window during which SLE of the corresponding layer is within acceptable bounds. Fig.\ref{fig:slel} shows an example of SLE monitoring for the GPU layer which is determined based on utilization, memory and temperature performance metrics. On anomaly detection, RCA is performed using a combination of statistical analysis (to rank the anomalous metrics) and few-shot learning with OpenAI’s Large Language Model. In Fig.\ref{fig:rca}, RCA identifies anomalous metrics in the application layer as symptoms and provides the root cause explanation comprising GPU utilization and network congestion issues with possible remediation steps, where one of the steps is to optimize workload distribution. Additionally, we utilize flow analysis to show GPU to GPU path tracing (a sample is shown in Fig.\ref{fig:uisample}) and the flow volume breakdown. This is helpful in seeing network related actionable insights in play, ex. load balancing. 

Our dashboard also provides actionable insights to address the root cause. In Fig.\ref{fig:rca}, one of the shown actionable insights is to optimize GPU workload distribution. Additionally, we utilize flow analysis to show GPU to GPU path tracing (a sample is shown in Fig \ref{fig:uisample}.) and the flow volume breakdown on per-queue pair basis. This could be helpful in seeing network related actionable insights in play, ex. load balancing.

\section{Conclusion}

In this demonstration, we present an end-to-end (E2E) assurance framework for distributed machine learning applications in the datacenter. We show a SaaS-based solution that offers comprehensive observability across various layers of the E2E path, including the application, graphical processing units (GPUs), and network fabric, by utilizing telemetry and log data. We show how it effectively troubleshoots workload performance issues by correlating multiple telemetry sources with the collective communication behavior of the workloads. An instrumented collective communication framework enables us to do this, which is crucial for building a dependency graph. Furthermore, this enhanced observability allows us to monitor cross-layer Service Level Expectations (SLE). If SLE degradation is detected through anomaly detection, an automated root cause analysis service can accurately identify the bottleneck and suggest remediation steps. Additionally, flow-level telemetry combined with log data provides insights into the packet path for GPU-to-GPU communication within the network. This demo captures the mentioned features to provide assurance for the mentioned applications.

\bibliographystyle{IEEEtran}
\bibliography{refs}

\begin{thebibliography}{1}
\providecommand{\url}[1]{#1}
\csname url@samestyle\endcsname
\providecommand{\newblock}{\relax}
\providecommand{\bibinfo}[2]{#2}
\providecommand{\BIBentrySTDinterwordspacing}{\spaceskip=0pt\relax}
\providecommand{\BIBentryALTinterwordstretchfactor}{4}
\providecommand{\BIBentryALTinterwordspacing}{\spaceskip=\fontdimen2\font plus
\BIBentryALTinterwordstretchfactor\fontdimen3\font minus \fontdimen4\font\relax}
\providecommand{\BIBforeignlanguage}[2]{{%
\expandafter\ifx\csname l@#1\endcsname\relax
\typeout{** WARNING: IEEEtran.bst: No hyphenation pattern has been}%
\typeout{** loaded for the language `#1'. Using the pattern for}%
\typeout{** the default language instead.}%
\else
\language=\csname l@#1\endcsname
\fi
#2}}
\providecommand{\BIBdecl}{\relax}
\BIBdecl

\bibitem{10.1145/3377454}
\BIBentryALTinterwordspacing
J.~Verbraeken, M.~Wolting, J.~Katzy, J.~Kloppenburg, T.~Verbelen, and J.~S. Rellermeyer, ``A survey on distributed machine learning,'' \emph{ACM Comput. Surv.}, vol.~53, no.~2, Mar. 2020. [Online]. Available: \url{https://doi.org/10.1145/3377454}
\BIBentrySTDinterwordspacing

\bibitem{10.1145/2934872.2934908}
\BIBentryALTinterwordspacing
C.~Guo, H.~Wu, Z.~Deng, G.~Soni, J.~Ye, J.~Padhye, and M.~Lipshteyn, ``Rdma over commodity ethernet at scale,'' in \emph{Proceedings of the 2016 ACM SIGCOMM Conference}, ser. SIGCOMM '16.\hskip 1em plus 0.5em minus 0.4em\relax New York, NY, USA: Association for Computing Machinery, 2016, p. 202–215. [Online]. Available: \url{https://doi.org/10.1145/2934872.2934908}
\BIBentrySTDinterwordspacing

\bibitem{weingram2023xccl}
A.~Weingram, Y.~Li, H.~Qi, D.~Ng, L.~Dai, and X.~Lu, ``xccl: A survey of industry-led collective communication libraries for deep learning,'' \emph{Journal of Computer Science and Technology}, vol.~38, no.~1, pp. 166--195.

\bibitem{torchmetrics}
\BIBentryALTinterwordspacing
N.~S. Detlefsen, J.~Borovec, J.~Schock, A.~Harsh, T.~Koker, L.~D. Liello, D.~Stancl, C.~Quan, M.~Grechkin, and W.~Falcon, ``Torchmetrics.'' [Online]. Available: \url{https://www.pytorchlightning.ai"}
\BIBentrySTDinterwordspacing

\bibitem{nvidia_dcgm}
N.~Corporation, ``Nvidia data center gpu manager (dcgm),'' \url{https://developer.nvidia.com/nvidia-data-center-gpu-manager}, 2023, accessed: 2025-01-06.

\end{thebibliography}

\end{document}